\documentclass[12pt]{article}
\begin{document}
\begin{titlepage}
\begin{flushright}
hep-th/0307076\\
TIT/HEP-502\\
July, 2003\\
\end{flushright}
\vspace{0.5cm}
\begin{center}
{\Large \bf Supersymmetric Gauge Theories\\
on Noncommutative Superspace
}
\lineskip .75em
\vskip2.5cm
{\large Takeo Araki},\
{\large Katsushi Ito} \
and {\large Akihisa Ohtsuka}
\vskip 2.5em
 {\large\it Department of Physics\\
Tokyo Institute of Technology\\
Tokyo, 152-8551, Japan}  \vskip 4.5em
\end{center}
\begin{abstract}
We study four dimensional supersymmetric gauge theory on the noncommutative 
superspace, recently proposed by Seiberg.
We construct the gauge-invariant action of ${\cal N}=1$ super Yang-Mills
theory with chiral and antichiral superfields, which has ${\cal 
N}={1\over2}$ supersymmetry on the noncommutative superspace.
We also construct the action of ${\cal N}=2$ super Yang-Mills theory.
It is shown that this theory has only ${\cal N}={1\over2}$ supersymmetry.
\end{abstract}
\end{titlepage}
\baselineskip=0.7cm
The deformation of superspace\cite{ncsuper,DeGrNi} has been attracted 
much attention recently.
In particular, Ooguri and Vafa studied the $C$-deformation of
${\cal N}=1$ supersymmetric gauge theories in four dimensions and 
computed the coupling to the graviphoton superfield arising from 
the higher genus amplitudes in (topological) superstring theory
\cite{OoVa}.
They introduced the noncommutativity only in the Grassmann odd 
coordinates,which breaks spacetime supersymmetry explicitly.
Recently, Seiberg\cite{Se} proposed another type of deformation which 
introduces noncommutativity both in Grassmann even and odd 
coordinates but imposes the commutativity in the chiral coordinates.
This deformation is shown to keep the ${\cal N}={1\over 2}$ 
supersymmetry and have some interesting properties in the field 
theoretical viewpoint.
Some recent papers deal with subjects related to this noncommutative 
superspace\cite{FeRe}.

 In the paper\cite{Se}, the deformation of four dimensional ${\cal N}=1$ 
supersymmetric
Yang-Mills theory has been studied.
In this paper, we study ${\cal N}=1$ supersymmetric gauge theory coupled 
with (anti-) chiral superfields on the noncommutative superspace.
We construct the gauge invariant Lagrangian in which the product of 
fields are defined by the star-product.
This theory has also ${\cal N}={1\over2}$ supersymmetry.
We apply this formulation to the case of ${\cal N}=2$ supersymmetric
Yang-Mills theory, where the chiral superfield belongs to the adjoint 
representation.
It is an interesting problem to examine whether the theory on the 
noncommutative superspace possesses further supersymmetry.
We will show, however, that only the original ${\cal N}={1\over2}$
supersymmetry is preserved in this formulation.

We begin with introducing noncommutative superspace as in \cite{Se}.
We follow the conventions of \cite{WeBa}.
Let $(x^{\mu},\theta^{\alpha},\bar{\theta}^{\dot{\alpha}})$ be the 
supercoordinates of superspace. Here $\mu=0,1,2,3$ and $\alpha , 
\dot{\alpha}=1,2$.
The Grassmann odd coordinates $\theta^{\alpha}$ obey the 
anticommutation relations
\begin{equation}
 \left\{ \theta^{\alpha},\theta^{\beta}\right\}=C^{\alpha\beta}.
\end{equation}
The product of functions of $\theta$ is Weyl ordered by using the 
star product, which is
the fermionic version of the Moyal product:
\begin{equation}
 f(\theta)*g(\theta)=f(\theta)
\exp\left(
- 
{C^{\alpha\beta}\over2}
\overleftarrow{\partial\over\partial\theta^{\alpha}}
\overrightarrow{\partial\over\partial\theta^{\beta}}
\right)g(\theta). 
\end{equation}
We assume that the fermionic coordinates 
$\bar{\theta}^{\dot{\alpha}}$ 
satisfy the ordinary (anti-)commutation relations
\begin{equation}
\left\{\bar{\theta}^{\dot{\alpha}},\bar{\theta}^{\dot{\beta}} \right\}
=\left\{ \bar{\theta}^{\dot{\alpha}},\theta^{\alpha}\right\}
=\left[ \bar{\theta}^{\dot{\alpha}},x^{\mu}\right]=0.
\end{equation}
We assume that  the chiral coordinates \begin{equation}
 y^{\mu}=x^{\mu}
+i\theta^{\alpha}\sigma^{\mu}_{\alpha\dot{\alpha}}
\bar{\theta}^{\dot{\alpha}}
\end{equation}
are commutative, {\it i.e.}
\begin{equation}
 \left[ y^{\mu},y^{\nu}\right]=[y^{\mu},\theta^{\alpha}]=
\left[y^{\mu},\bar{\theta}^{\dot{\alpha}}\right]=0,
\end{equation}
instead of requiring the commutativity of the spacetime coordinates
$x^{\mu}$.
These relations imply that $x^{\mu}$ are noncommutative
\begin{equation}
 \left[ x^{\mu},x^{\nu}\right]=\bar{\theta}\bar{\theta} 
 C^{\mu\nu},\quad
[x^{\mu},\theta^{\alpha}]=iC^{\alpha\beta}
\sigma^{\mu}_{\beta\dot{\beta}}
\bar{\theta}^{\dot{\beta}},
\end{equation}
where
\begin{equation}
 C^{\mu\nu}=C^{\alpha\beta}\epsilon_{\beta\gamma}
(\sigma^{\mu\nu})_{\alpha} {}^{\gamma}.
\end{equation}
On this noncommutative superspace, the supercovariant derivatives and
the supercharges are
defined by
\begin{equation}
 D_{\alpha}={\partial\over \partial \theta^{\alpha}}
+2i \sigma^{\mu}_{\alpha\dot{\alpha}}\bar{\theta}^{\dot{\alpha}}
{\partial \over \partial y^{\mu}},\quad
\bar{D}_{\dot{\alpha}}=-{\partial\over\partial
\bar{\theta}^{\dot{\alpha}}},
\end{equation}
and
\begin{equation}
 Q_{\alpha}={\partial\over \partial \theta^{\alpha}},\quad
\bar{Q}_{\dot{\alpha}}=-
{\partial\over \partial\bar{\theta}^{\dot{\alpha}}}
+2i\theta^{\alpha}
\sigma^{\mu}_{\alpha\dot{\alpha}}{\partial \over \partial y^{\mu}},
\end{equation}
respectively.

 The chiral superfield satisfying $\bar{D}_{\dot{\alpha}}\Phi=0$
is expressed in terms of component fields
\begin{equation}
\Phi(y,\theta)=A(y)+\sqrt{2}\theta\psi(y)+\theta\theta F(y).
\end{equation}
While the antichiral superfield $\bar{\Phi}(\bar{y},\bar{\theta})$ 
with
$\bar{y}^{\mu}=y^{\mu}-2i\theta^{\alpha}
\sigma^{\mu}_{\alpha\dot{\alpha}}
\bar{\theta}^{\dot{\alpha}}$ can be written in  the Weyl ordered form:
\begin{eqnarray}
 \bar{\Phi}(\bar{y},\bar{\theta})
&=& \bar{A}(\bar{y})+\sqrt{2}\bar{\theta}\bar{\psi}(\bar{y})
-2i\theta \sigma^{\mu}\bar{\theta} \partial_{\mu}\bar{A}(\bar{y})
\nonumber\\
&&+\bar{\theta}\bar{\theta}
\left( \bar{F}(\bar{y})
+i\sqrt{2}\theta\sigma^{\mu}\partial_{\mu}\bar{\psi}(\bar{y})
+\theta\theta\partial_{\mu}\partial^{\mu}\bar{A}(\bar{y})\right).
\label{eq:def barphi}
\end{eqnarray}

We next introduce the vector superfield $V$ in the certain
matrix representation of the gauge group.
We take the basis ${t^a}$ of the gauge group satisfying 
${\rm tr}(t^a t^b) 
=k\delta^{ab}$ and $[t^a,t^b]=i t^{abc}t^c$.
Define $e^{V}$ by $\sum_{n=0}^{\infty}{1\over n!}(V^n)_{*}$, where
$(V^n)_{*}$ is the $n$-th power of $V$ defined by using the 
star-product.
$e^{V}$ transforms as $e^{V}\rightarrow
e^{V'}=e^{-i\bar{\Lambda}}*
e^{V}*e^{i\Lambda}$, or infinitesimally
\begin{equation}
 \delta e^{V}=-i\bar{\Lambda}*e^{V}+i e^{V}*\Lambda. \label{eq:gauge1}
\end{equation}
Here $\Lambda$ and $\bar{\Lambda}$ are matrices of chiral and
antichiral superfields respectively.
The vector superfield $V$ in the Wess-Zumino gauge is
\begin{eqnarray}
 V(y,\theta,\bar{\theta})&=&
-\theta\sigma^{\mu}\bar{\theta} A_{\mu}(y)
+i\theta\theta\bar{\theta}\bar{\lambda}(y)
-i\bar{\theta}\bar{\theta}\theta^{\alpha}
\left(\lambda_{\alpha}(y)
+{1\over4}\epsilon_{\alpha\beta}C^{\beta\gamma}
\sigma^{\mu}_{\gamma\dot{\gamma}}\left\{
\bar{\lambda}^{\dot{\gamma}},A_{\mu}
\right\}(y)
\right)
\nonumber\\
&&+{1\over2} \theta\theta\bar{\theta}\bar{\theta}\left( D(y)
-i\partial_{\mu}A^{\mu}(y)\right).
\label{eq:vect}
\end{eqnarray}
The $C$-deformed part in $V$  is introduced such 
that the component fields 
transform canonically under the
gauge transformation\cite{Se}: in terms of the component fields, 
(\ref{eq:gauge1}) becomes \begin{eqnarray}
 \delta A_{\mu}&=&-2\partial_{\mu}\varphi+i[\varphi,A_{\mu}], 
 \nonumber\\
\delta \bar{\lambda}&=&i [\varphi,\bar{\lambda}], \nonumber\\
\delta\lambda&=& i[\varphi,\lambda], \nonumber\\
\delta D&=&i[\varphi,D]. \label{eq:gau1}
\end{eqnarray}
Then the gauge transformation which preserves 
the gauge (\ref{eq:vect}) is given by
\begin{eqnarray}
\Lambda(y,\theta)&=& -\varphi(y),\nonumber\\
\bar{\Lambda}(\bar{y},\bar{\theta})&=&
-\varphi(\bar{y})
-{i\over2}\bar{\theta}\bar{\theta}C^{\mu\nu}
\{ \partial_{\mu}\varphi, 
A_{\nu}\}(\bar{y}).
\end{eqnarray}
The chiral and antichiral field strengths are given by
\begin{eqnarray}
 W_{\alpha}&=&-{1\over 4} \bar{D}\bar{D} e^{-V} D_{\alpha} e^V,
\nonumber\\
\overline{W}_{\dot{\alpha}}&=&{1\over 4} DD e^{V} \bar{D}_{\dot{\alpha}} 
e^{-V}, \end{eqnarray}
which are also defined by the star-product.

We now consider ${\cal N}=1$ supersymmetric gauge theory with a chiral
superfield $\Phi$ in the fundamental representation and
an antichiral superfield $\bar{\Phi}$ in the antifundamental
representation.
Since the $C$-deformed part in (\ref{eq:vect}) does not take value 
in the Lie algebra of the gauge group, we will consider 
the $U(N)$ gauge group for simplicity.
Under the gauge transformation, $\Phi$ and $\bar{\Phi}$ transform as
\begin{equation}
\Phi\rightarrow e^{-i\Lambda}*\Phi,\quad
\bar{\Phi}\rightarrow \bar{\Phi}*e^{i\bar{\Lambda}}.
\end{equation}
Their infinitesimal forms are
\begin{equation}
 \delta \Phi=-i \Lambda*\Phi,\quad
\delta \bar{\Phi}=i \bar{\Phi}*\bar{\Lambda}.
\label{eq:gauge2}
\end{equation}
We express these transformations in terms of the component fields.
For a chiral superfield $\Phi$, the gauge transformation is the same
as the commutative one: \begin{eqnarray}
\delta A(y)&=& i \varphi A(y), \quad
\delta \psi(y)= i\varphi\psi(y),\quad \delta F(y)=  i\varphi F(y). 
\label{eq:gau2}
\end{eqnarray}
For an antichiral superfield $\bar{\Phi}$, if we consider the usual 
component fields in eq.(\ref{eq:def barphi}), 
the gauge transformation rule 
will be changed due to the $C$-dependent term in $\bar{\Lambda}$. 
In fact, we have 
\begin{eqnarray}
\delta \bar{\Phi} (\bar{y}, \bar{\theta})
&=& -i \bar{A} \varphi(\bar{y}) + \sqrt{2} 
\bar{\theta}_{\dot{\alpha}} (-i \bar{\psi}^{\dot{\alpha}}  
\varphi (\bar{y})) \nonumber\\
& & + \bar{\theta} \bar{\theta} \left( 
-i\bar{F} \varphi (\bar{y}) 
-2iC^{\mu \nu} \partial_{\mu} \bar{A}  
\partial_{\nu} \varphi (\bar{y}) 
+ \frac{1}{2} C^{\mu \nu} \bar{A} 
\{ \partial_{\mu}\varphi , A_{\nu} \}(\bar{y}) \right), 
\label{eq:DeltaBarPhi}
\end{eqnarray}
where we have used the formula:
\begin{eqnarray*}
f(\bar{y}) \ast g(\bar{y}) &=& 
f(\bar{y}) \exp{\left( 2\bar{\theta}\bar{\theta} C^{\mu\nu} 
\overleftarrow{\frac{\partial}{\partial \bar{y}^{\mu}}} 
\overrightarrow{\frac{\partial}{\partial \bar{y}^{\nu}}} \right)} 
g(\bar{y}) .
\end{eqnarray*}
We note that the $\bar{\theta}\bar{\theta}$-term 
in (\ref{eq:DeltaBarPhi}) is expressed as 
\begin{equation}
-i \left( \bar{F} -iC^{\mu\nu} \partial_{\mu} (\bar{A} A_{\nu}) 
+ \frac{1}{4} C^{\mu\nu} \bar{A} A_{\mu} A_{\nu} \right) \varphi 
+ \delta \left( iC^{\mu\nu} \partial_{\mu} (\bar{A} A_{\nu}) 
- \frac{1}{4} C^{\mu\nu} \bar{A} A_{\mu} A_{\nu} \right). 
\end{equation}
Therefore if we define the antichiral superfield as 
\begin{eqnarray}
\bar{\Phi}(\bar{y}, \bar{\theta}) 
= \bar{A} (\bar{y}) + \sqrt{2} \bar{\theta} \bar{\psi}(\bar{y})
+ \bar{\theta} \bar{\theta} \left( \bar{F} (\bar{y}) 
+ iC^{\mu\nu} \partial_{\mu} (\bar{A} A_{\nu}) (\bar{y}) 
- \frac{1}{4} C^{\mu\nu} \bar{A} A_{\mu} A_{\nu} (\bar{y})
 \right), \label{eq:new barphi}
\end{eqnarray}
the component fields are shown to transform canonically: 
\begin{eqnarray}
\delta \bar{A}(y) = -i  \bar{A} \varphi(y), \quad \delta \bar{\psi}(y)
= - i\bar{\psi} \varphi(y), \quad \delta \bar{F}(y) 
=  -i \bar{F}\varphi(y). 
\label{eq:gau3}
\end{eqnarray}

The gauge invariant Lagrangian is given by  
\begin{equation}
 {\cal L}=\int d^2\theta d^2 \theta \bar{\Phi}*e^{V}*\Phi
+{1\over 16 kg^2}\left(\int d^2\theta {\rm tr}W^{\alpha}*W_{\alpha}
+\int d^2\bar{\theta} {\rm tr} \overline{W}_{\dot{\alpha}}*
\overline{W}^{\dot{\alpha}}\right), 
\label{eq:lag1}
\end{equation}
which is not affected by the redefinition of $\bar{F}$. 
The $F$-terms have 
been computed in \cite{Se}: \begin{eqnarray}
\left.{\rm tr} W^{\alpha}W_{\alpha}\right|_{\theta\theta}
&=& \left.{\rm tr} W^{\alpha}W_{\alpha}(C=0)\right|_{\theta\theta}
-i C^{\mu\nu}{\rm tr}F_{\mu\nu}\bar{\lambda}\bar{\lambda}
+{|C|^2\over4}{\rm tr}(\bar{\lambda}\bar{\lambda})^2, \nonumber\\
\left.{\rm tr} 
\bar{W}_{\dot{\alpha}}\bar{W}^{\dot{\alpha}}\right|_{\bar{\theta}
\bar{\theta}}
&=& \left.{\rm tr} \bar{W}_{\dot{\alpha}}\bar{W}^{\dot{\alpha}}(C=0)
\right|_{\bar{\theta}\bar{\theta}}
-i C^{\mu\nu}{\rm tr}F_{\mu\nu}\bar{\lambda}\bar{\lambda}
+{|C|^2\over4}{\rm tr}(\bar{\lambda}\bar{\lambda})^2 \nonumber\\
&&+\mbox{total derivative}. \end{eqnarray}
Here $|C|^2=C^{\mu\nu}C_{\mu\nu}$
and \begin{eqnarray}
 \left.{\rm tr} W^{\alpha}W_{\alpha}(C=0)\right|_{\theta\theta}&=& 
 {\rm tr} 
\left( -2i \bar{\lambda} \bar{\sigma}^{\mu} {\cal D}_{\mu} \lambda - 
\frac{1}{2} F^{\mu \nu} F_{\mu \nu} + D^2 + \frac{i}{4} F^{\mu \nu} 
F^{\rho \sigma} \varepsilon_{\mu \nu \rho \sigma} \right), \nonumber\\
 \left.{\rm tr} \bar{W}_{\dot{\alpha}}\bar{W}^{\dot{\alpha}}(C=0)
\right|_{\bar{\theta}\bar{\theta}}&=& {\rm tr} \left(-2i\bar{\lambda}
\bar{\sigma}^{\mu} {\cal D}_{\mu} \lambda -\frac{1}{2} F^{\mu \nu} 
F_{\mu \nu} + D^2 - \frac{i}{4} F^{\mu \nu} F^{\rho \sigma} 
\varepsilon_{\mu \nu \rho \sigma} \right), \end{eqnarray}
where 
\begin{eqnarray}
 F_{\mu\nu}&=& \partial_{\mu}A_{\nu}-\partial_{\nu}A_{\mu}+{i\over2}
[A_{\mu},A_{\nu}], \nonumber\\
{\cal D}_{\mu}\lambda&=& \partial_{\mu}\lambda+{i\over2}[A_{\mu},
\lambda].
\end{eqnarray}
The gauge invariance of the $D$-term is ensured by the transformation 
properties of the superfields (\ref{eq:gauge1}) and (\ref{eq:gauge2}).
 The $D$-term in the Wess-Zumino gauge can be computed by using
\begin{eqnarray}
 V^2&=& \bar{\theta}\bar{\theta}
\left[
-{1\over2}\theta\theta A_{\mu}A^{\mu}
-{1\over2} C^{\mu\nu}A_{\mu}A_{\nu}
+{i\over2}\theta_{\alpha}C^{\alpha\beta}
\sigma^{\mu}_{\beta\dot{\alpha}}
[A_{\mu}, \bar{\lambda}^{\dot{\alpha}}]-{1\over 8} |C|^2 
\bar{\lambda}\bar{\lambda}\right], \nonumber\\
V^3&=&0.
\end{eqnarray}
{}From these formulas, $\bar{\Phi}*e^{V}*\Phi$
becomes 
\begin{eqnarray}
\bar{\Phi} \ast e^{V} \ast \Phi &=& \bar{\Phi} 
\ast (1+V+\frac{1}{2}V^2) \ast \Phi. 
\label{eq:pb(1+v+v^2)p}
\end{eqnarray}
The $\theta\theta\bar{\theta}\bar{\theta}$ component of 
each term in the r.h.s. of (\ref{eq:pb(1+v+v^2)p}) is given by 
\begin{eqnarray}
\bar{\Phi} \ast \Phi |_{\theta \theta \bar{\theta} \bar{\theta}} &=& 
\bar{F} F +i C^{\mu \nu} \partial_{\mu} (\bar{A} A_{\nu}) F
-\frac{1}{4} C^{\mu \nu} \bar{A} A_{\mu} A_{\nu} F
+ i \sigma^{\mu}_{\alpha \dot{\alpha}} \partial_{\mu} 
\bar{\psi}^{\dot{\alpha}} \psi^{\alpha} + \partial^2 \bar{A} A , \\
\bar{\Phi} \ast V \ast {\Phi}|_{\theta \theta \bar{\theta} 
\bar{\theta}} 
&=& + i \frac{\sqrt{2}}{2} \bar{A} \lambda \psi - i\frac{\sqrt{2}}{2} 
\bar{\psi} \bar{\lambda} A + \frac{1}{2} \bar{\psi}^{\dot{\alpha}} 
\sigma^{\mu}_{\alpha \dot{\alpha}} A_{\mu} \psi^{\alpha}
+ \frac{1}{2} \bar{A} (D-i\partial_{\mu} A^{\mu}) A \nonumber\\
& &  - i  \partial^{\mu} \bar{A} A_{\mu} A
+ i \frac{\sqrt{2}}{8} C^{\beta \gamma} 
\sigma^{\mu}_{\gamma \dot{\gamma}} 
\bar{A} \{ \bar{\lambda}^{\dot{\gamma}}, A_{\mu} \}  \psi_{\beta} 
\nonumber\\
& & - i C^{\mu \nu} \partial_{\mu} \bar{A} A_{\nu} F
+ \frac{\sqrt{2}}{2} \epsilon_{\beta \alpha} C^{\beta \gamma} 
\bar{\sigma}^{\mu \dot{\alpha} \alpha} \partial_{\mu} \bar{A} 
\bar{\lambda}_{\dot{\alpha}} \psi_{\gamma},  \\
\bar{\Phi} \ast V^2 \ast \Phi |_{\theta \theta \bar{\theta} 
\bar{\theta}} 
&=& -\frac{1}{2} \bar{A} A_{\mu} A^{\mu} A - \frac{1}{2} C^{\mu \nu} 
\bar{A} A_{\mu} A_{\nu} F +i \frac{\sqrt{2}}{4} C^{\alpha \beta} 
\sigma^{\mu}_{\alpha \dot{\alpha}} \bar{A} [A_{\mu} , 
\bar{\lambda}^{\dot{\alpha}} ] \psi_{\beta} \nonumber\\
& &  - \frac{1}{8} |C|^2 \bar{A} \bar{\lambda} \bar{\lambda} F, 
\end{eqnarray}
up to total derivatives.  Thus we find 
\footnote{ 
If we take the ${\bar{\theta}\bar{\theta}}$ 
component of ${\bar{\Phi}}$ 
to be the same as eq.(\ref{eq:def barphi}) instead of 
eq.(\ref{eq:new barphi}), in eq.(\ref{eq:phieVphi}) \
$\frac{i}{2}C^{\mu\nu}\bar{A}F_{\mu\nu}F$ 
is replaced by 
$-i C^{\mu\nu}\partial_{\mu}\bar{A} A_{\nu}F 
- {1\over4} \bar{A}C^{\mu\nu}A_{\mu}A_{\nu}F$ . 
These terms seem to break the gauge invariance, 
but using the modified gauge transformation (\ref{eq:DeltaBarPhi}), 
the Lagrangian is shown to be gauge invariant. 
}
\begin{eqnarray}
\left. \bar{\Phi}*e^{V}*\Phi\right|_{\theta\theta\bar{\theta}
\bar{\theta}}
&=& \left.\bar{\Phi}e^{V}\Phi(C=0)
\right|_{\theta\theta\bar{\theta}\bar{\theta}} \nonumber\\
&&{} + \frac{i}{2} C^{\mu\nu} \bar{A} F_{\mu\nu} F 
-{1\over16} |C|^2\bar{A} \bar{\lambda}\bar{\lambda}F
-{\sqrt{2}\over2}
C^{\alpha\beta}({\cal D}_{\mu}\bar{A})\sigma^{\mu}_{\beta\dot{\alpha}}
\bar{\lambda}^{\dot{\alpha}} \psi_{\alpha},
\label{eq:phieVphi}
\end{eqnarray}
where
\begin{eqnarray}
 \left.\bar{\Phi}e^{V}\Phi
(C=0)\right|_{\theta\theta\bar{\theta}\bar{\theta}}
&=& \bar{F} F -i \bar{\psi} \bar{\sigma}^{\mu} {\cal D}_{\mu} \psi
-{\cal D}_{\mu}\bar{A} {\cal D}^{\mu}A + \frac{1}{2} \bar{A} D A + 
\frac{i}{\sqrt{2}} (\bar{A} \lambda \psi - \bar{\psi} \bar{\lambda} A)
\nonumber \\
\end{eqnarray}
and 
\begin{equation}
 {\cal D}_{\mu}\psi=\partial_{\mu}\psi+{i\over2}A_{\mu}\psi,\quad
{\cal D}_{\mu}A=\partial_{\mu}A+{i\over2}A_{\mu}A.
\end{equation}
To summarize, the total Lagrangian (\ref{eq:lag1}) becomes 
\begin{eqnarray}
{\cal L}&=&
\frac{1}{ 16kg^2}{\rm tr}\left(
-4i 
\bar{\lambda} \bar{\sigma}^{\mu} {\cal D}_{\mu} \lambda - F^{\mu 
\nu} F_{\mu \nu} + 2 D^2 
\right)\nonumber\\
&&{}+\bar{F} F -i \bar{\psi} \bar{\sigma}^{\mu} {\cal D}_{\mu} 
\psi-{\cal D}_{\mu}\bar{A} {\cal D}^{\mu}A + \frac{1}{2} \bar{A} D A 
+\frac{i}{\sqrt{2}}(\bar{A} \lambda \psi- \bar{\psi} \bar{\lambda} A)
\nonumber\\
& &{} + \frac{1}{16kg^2} {\rm tr} \left( -2i C^{\mu \nu} F_{\mu \nu} 
\bar{\lambda} \bar{\lambda} 
+ \frac{|C|^2}{2} (\bar{\lambda} \bar{\lambda})^2 \right) \nonumber\\
&&{} + \frac{i}{2} C^{\mu\nu} \bar{A} F_{\mu\nu} F 
-\frac{\sqrt{2}}{2}C^{\alpha \beta}\sigma^{\mu}_{\alpha \dot{\alpha}}
{\cal D}_{\mu} \bar{A} \bar{\lambda}^{\dot{\alpha}} \psi_{\beta} 
- \frac{|C|^2}{16} \bar{A} \bar{\lambda} \bar{\lambda} F. 
\label{eq:lag2}
\end{eqnarray}

We next discuss the supersymmetry of the theory.
We expect that the supercharge $Q_{\alpha}$ is conserved.
In the case of $C=0$, the supersymmetry transformation is given by
\begin{eqnarray}
\delta^{0}_{\xi}A&=& \sqrt{2}\xi\psi, \quad
\delta^{0}_{\xi}\bar{A}=0, 
\nonumber\\
\delta^{0}_{\xi}\psi_{\alpha}&=&
\sqrt{2}\xi_{\alpha} F,\quad 
\delta^{0}_{\xi}\bar{\psi}_{\dot{\alpha}}
=-i\sqrt{2}{\cal D}_{\mu}\bar{A}(\xi\sigma^{\mu})_{\dot{\alpha}}, 
\nonumber\\
\delta^{0}_{\xi}F&=&0,\quad
\delta^{0}_{\xi}\bar{F}=
-i\sqrt{2}{\cal D}_{\mu}\bar{\psi}\bar{\sigma}^{\mu} \xi
-i \bar{A}\xi\lambda, 
\nonumber\\
\delta^{0}_{\xi}A_{\mu}&=&
-i\bar{\lambda}\bar{\sigma}_{\mu}\xi, 
\nonumber\\
\delta^{0}_{\xi}\lambda_{\alpha}&=& i\xi_{\alpha} D+
(\sigma^{\mu\nu}\xi)_{\alpha}F_{\mu\nu}
,\quad
\delta^{0}_{\xi}\bar{\lambda}_{\dot{\alpha}}=0, 
\nonumber\\
\delta^{0}_{\xi}D&=& -\xi \sigma^{\mu}{\cal D}_{\mu}\bar{\lambda}.
\label{eq:susy0}
\end{eqnarray}
We would like to find the $C$-dependent transformation
\begin{equation}
 \delta_{\xi}=\delta^{0}_{\xi}+\delta^{C}_{\xi}
\end{equation}
which leaves the Lagrangian invariant up to total derivatives. 
$\delta^{C}_{\xi}\lambda$ is obtained by acting $Q_{\alpha}$ 
on $V$
\begin{equation}
 \delta^{C}_{\xi}\lambda_{\alpha}
={i\over2} (\sigma^{\mu\nu}\xi)_{\alpha} C_{\mu\nu} 
\bar{\lambda}\bar{\lambda}.
\label{eq:C-deform lambda}
\end{equation}
This transformation leaves 
the $C$-deformed $F$-term invariant\cite{Se}. 
The $C$-dependent terms arising from the $D$-term 
in the Lagrangian is
\begin{eqnarray}
\Delta {\cal L}  &\equiv& \frac{i}{2} C^{\mu\nu} \bar{A} F_{\mu\nu}F 
- \frac{\sqrt{2}}{2} C^{\alpha \beta} 
\sigma^{\mu}_{\alpha \dot{\alpha}} 
{\cal D}_{\mu} \bar{A} \bar{\lambda}^{\dot{\alpha}} \psi_{\beta} 
- \frac{|C|^2}{16} \bar{A} \bar{\lambda} \bar{\lambda} F .
\end{eqnarray}
The change of $\Delta{\cal L}$ under the transformation
(\ref{eq:susy0}) is given by
\begin{eqnarray}
 \delta^{0}_{\xi}\Delta{\cal L}
&=& -C^{\mu\nu}\bar{A} \xi\sigma_{\nu} D_{\mu}\bar{\lambda}F
+\frac{\sqrt{2}}{4}C^{\mu\nu} (\sigma_{\mu\nu} \xi)^{\alpha}
\bar{A} \bar{\lambda}\bar{\lambda}\psi_{\alpha}
- C^{\mu\nu}D_{\mu}\bar{A}
\xi\sigma_{\nu} \bar{\lambda}F . 
\end{eqnarray}
The second term is cancelled by the term 
coming from $\delta_{\xi}^C \Delta{\cal L}$ as a consequence of 
eq.(\ref{eq:C-deform lambda}): 
\begin{eqnarray}
i \frac{\sqrt{2}}{2} \bar{A} (\delta_{\xi}^C\lambda^{\alpha}) 
\psi_{\alpha} &=& -\frac{\sqrt{2}}{4} C^{\mu\nu}
(\sigma_{\mu\nu})^{\alpha} \bar{A} \bar{\lambda}\bar{\lambda}
\psi_{\alpha}.
\end{eqnarray}
The first and third terms can be removed 
by adding a $C$-dependent term 
\footnote{If we adopt the usual definition of $\bar{F}$, 
the right hand side of (\ref{eq:Fbar trans.}) 
is replaced with $\frac{i}{4} C^{\mu\nu} 
\bar{A} \{ A_{\mu} , \bar{\lambda} \sigma_{\nu} \} \xi$.}
to the ordinary 
supersymmetry transformation 
of $\bar{F}$ : 
\begin{equation}
 \delta^{C}_{\xi}\bar{F}=C^{\mu\nu} \left\{ \partial_{\mu} \left( 
\bar{A} \xi \sigma_{\nu} \bar{\lambda} \right) 
-\frac{i}{2} \left( \bar{A} \xi \sigma_{\nu} \bar{\lambda} \right) 
A_{\mu} \right\} .
\label{eq:Fbar trans.}
\end{equation}
Therefore, we have found   
${\cal N}=\frac{1}{2}$ supersymmetry transformation 
by deforming the $\bar{F}$ transformation, 
under which the Lagrangian (\ref{eq:lag2}) is invariant: 
\begin{eqnarray}
 \delta_{\xi}A&=& \sqrt{2}\xi\psi,\quad
\delta_{\xi}\bar{A}=0, 
\nonumber\\
\delta_{\xi}\psi_{\alpha}&=&
\sqrt{2}\xi_{\alpha} F,\quad
\delta_{\xi}\bar{\psi}_{\dot{\alpha}}
=-i\sqrt{2}{\cal D}_{\mu}\bar{A}(\xi\sigma^{\mu})_{\dot{\alpha}}, 
\nonumber\\
\delta_{\xi}F&=&0,\quad
\delta_{\xi}\bar{F}=
-i\sqrt{2}{\cal D}_{\mu}\bar{\psi}\bar{\sigma}^{\mu} \xi
-i \bar{A}\xi\lambda
+C^{\mu\nu} \left\{ \partial_{\mu} \left( 
\bar{A} \xi \sigma_{\nu} \bar{\lambda} \right) 
-\frac{i}{2} \left( \bar{A} \xi \sigma_{\nu} \bar{\lambda} \right) 
A_{\mu} \right\} , \nonumber\\
\delta_{\xi}A_{\mu}&=&
-i\bar{\lambda}\bar{\sigma}_{\mu}\xi, 
\nonumber\\
\delta_{\xi}\lambda_{\alpha}&=& i\xi_{\alpha} D+
(\sigma^{\mu\nu}\xi)_{\alpha}\left(F_{\mu\nu}+{i\over2} C_{\mu\nu} 
\bar{\lambda}\bar{\lambda}\right)
,\quad
\delta_{\xi}\bar{\lambda}_{\dot{\alpha}}=0, 
\nonumber\\
\delta_{\xi}D&=& -\xi \sigma^{\mu}{\cal D}_{\mu}\bar{\lambda}.
\end{eqnarray}
Another supersymmetry generated by $\bar{Q}_{\dot{\alpha}}$ is broken
explicitly as in \cite{Se}.

So far we have considered the case that $\Phi$ 
belongs to the fundamental
representation of the gauge group.
When we consider the case that $\Phi$ belongs to the adjoint
representation, we obtain ${\cal N}=2$ supersymmetric gauge theory. 
In this case, $\Phi$ and $\bar{\Phi}$ transform as 
\begin{eqnarray}
\Phi \to e^{-i\Lambda} \ast \Phi \ast e^{i\Lambda} ,\quad
\bar{\Phi} \to e^{-i\bar{\Lambda}} \ast \bar{\Phi} \ast 
e^{i\bar{\Lambda}} ,
\end{eqnarray}
where $\bar{\Phi}$ is defined by
\begin{eqnarray}
\bar{\Phi} &=& 
\bar{A}(\bar{y}) + \sqrt{2} \bar{\theta} \bar{\psi} (\bar{y}) 
\nonumber\\
& & + \bar{\theta}\bar{\theta} \left( \bar{F}(\bar{y}) 
+iC^{\mu\nu} \partial_{\mu} \{ \bar{A},A_{\nu} \} (\bar{y})
-\frac{g}{2}C^{\mu\nu} [A_{\mu}, \{ A_{\nu},\bar{A} \} ](\bar{y}) 
\right) ,
\end{eqnarray}
so that $\bar{A}$, $\bar{\psi}$ and $\bar{F}$ transform canonically.
The gauge invariant Lagrangian of ${\cal N}=2$ supersymmetric theory 
is 
\begin{equation}
 {\cal L}=\frac{1}{k} \int d^2\theta d^2 \theta {\rm tr} \left( 
 \bar{\Phi}*e^{V}*\Phi \ast e^{-V} \right)
+{1\over 16 kg^2}\left(\int d^2\theta {\rm tr}W^{\alpha}*W_{\alpha}
+\int d^2\bar{\theta} {\rm tr} \overline{W}_{\dot{\alpha}}*
\overline{W}^{\dot{\alpha}}\right) . 
\end{equation}
The $F$-term is the same as ${\cal N}=1$ case, and the $D$-term 
is given by 
\begin{eqnarray}
&&  {\rm tr} \left. \left( \bar{\Phi} \ast e^V \ast \Phi \ast e^{-V} 
\right) \right|_{\theta\theta\bar{\theta}\bar{\theta}} \nonumber\\
&&= {\rm tr} \left[ \bar{F}F -\frac{1}{2} [\bar{A},A] D
-i\bar{\psi}\bar{\sigma}^{\mu} {\cal D}_{\mu} \psi 
-{\cal D}_{\mu} \bar{A} {\cal D}^{\mu} A 
+i \frac{\sqrt{2}}{2} \left( \bar{\lambda} [A,\bar{\psi}] 
+[\bar{A},\psi] \lambda \right) \right] \nonumber\\
&&{}\quad  + {\rm tr} \left[ \frac{i}{2} C^{\mu\nu} F_{\mu\nu} 
\{ \bar{A},F \} 
-\frac{\sqrt{2}}{2}C^{\alpha\beta}\sigma^{\mu}_{\alpha\dot{\alpha}} 
\{ {\cal D}_{\mu} \bar{A} , \bar{\lambda}^{\dot{\alpha}} \} 
\psi_{\beta} 
-\frac{|C|^2}{16} \{ \bar{A},\bar{\lambda}_{\dot{\alpha}} \} [
\bar{\lambda}^{\dot{\alpha}},F] \right] ,
\end{eqnarray}
up to total derivatives. 

Rescaling $V$ to $2gV$ and $C^{\alpha\beta}$ to 
$\frac{1}{2g}C^{\alpha\beta}$ ,
the Lagrangian for ${\cal N}=2$ supersymmetric Yang-Mills theory 
becomes ${\cal L}^{{\cal N}=2}={\cal L}^{{\cal N}=2}(C=0)
+{\cal L}^{{\cal N}=2}_C$, where
\begin{eqnarray}
 {\cal L}^{{\cal N}=2}(C=0)&=&
\frac{1}{k}{\rm tr}\Bigl(
-\frac{1}{4}F^{\mu \nu}F_{\mu \nu}
-i \bar{\lambda}\bar{\sigma}^{\mu}{\cal D}_{\mu}\lambda
+\frac{1}{2}D^2 -({\cal D}^{\mu}\bar{A}){\cal D}_{\mu}A
-i \bar{\psi}\bar{\sigma}^{\mu}{\cal D}_{\mu}\psi
+\bar{F} F
\nonumber\\
&&
-i\sqrt{2} g[\bar{A},\psi]\lambda
+i\sqrt{2} g[A,\bar{\psi}]\bar{\lambda}
+gD[A,\bar{A}]
\Bigr), \\
{\cal L}^{{\cal N}=2}_C &=&
\frac{1}{k} {\rm tr} 
\left( - \frac{i}{2} C^{\mu \nu} F_{\mu \nu} \bar{\lambda} 
\bar{\lambda} 
+ \frac{1}{8} |C|^2 (\bar{\lambda} \bar{\lambda})^2 
+\frac{i}{2} C^{\mu\nu} F_{\mu\nu} \{ \bar{A},F \} \right. \nonumber\\
& & \left. 
- \frac{\sqrt{2}}{2} C^{\alpha \beta} \{ {\cal D}_{\mu} \bar{A} , 
(\sigma^{\mu} \bar{\lambda})_{\alpha} \} \psi_{\beta}  
- \frac{1}{16} |C|^2 \{ \bar{A} , \bar{\lambda}\} [\bar{\lambda} , F] 
\right). 
\end{eqnarray}
${\cal L}^{{\cal N}=2}(C=0)$ is invariant under ${\cal N}=2$ 
supersymmetry transformations:
\begin{eqnarray}
 \delta_{\xi}A&=& \sqrt{2}\xi\psi,\quad
\delta_{\xi}\bar{A}=\sqrt{2}\bar{\xi}\bar{\psi}, 
\nonumber\\
\delta_{\xi}\psi&=&i\sqrt{2}\sigma^{\mu}\bar{\xi}{\cal D}_{\mu}A
+\sqrt{2}\xi F, 
\quad
\delta_{\xi}\bar{\psi}=
-i \sqrt{2} {\cal D}_{\mu}\bar{A}\xi\sigma^{\mu}
+\sqrt{2}\bar{F}\bar{\xi}, 
\nonumber\\
\delta_{\xi}F&=&
i\sqrt{2}\bar{\xi}\bar{\sigma}^{\mu}{\cal D}_{\mu}\psi
- 2ig [A,\bar{\xi}\bar{\lambda}],
\quad
\delta_{\xi}\bar{F}=-i\sqrt{2}{\cal D}_{\mu}\bar{\psi}
\bar{\sigma}^{\mu}\xi
- 2ig [\xi\lambda ,\bar{A}], 
\nonumber\\
\delta_{\xi}A_{\mu}&=&
-i\bar{\lambda} \bar{\sigma}_{\mu} \xi
+i\bar{\xi}\bar{\sigma}_{\mu}\lambda, 
\nonumber\\
\delta_{\xi}\lambda&=&
\sigma^{\mu \nu}\xi F_{\mu \nu}+i\xi D, 
\quad \delta_{\xi}\bar{\lambda}=
-\bar{\xi} \bar{\sigma}^{\mu \nu} F_{\mu \nu}-i\bar{\xi} D, 
\nonumber\\
\delta_{\xi}D&=& -\xi \sigma^{\mu}{\cal D}_{\mu}\bar{\lambda}
-{\cal D}_{\mu}\lambda\sigma^{\mu}\bar{\xi}, 
\end{eqnarray}
\begin{eqnarray}
 \delta_{\eta}A&=& \sqrt{2}\eta\lambda,\quad
\delta_{\eta}\bar{A}=\sqrt{2}\bar{\eta}\bar{\lambda}, 
\nonumber\\
\delta_{\eta}\lambda&=&i\sqrt{2}\sigma^{\mu}\bar{\eta}{\cal D}_{\mu}A
+\sqrt{2}\eta F, 
\quad
\delta_{\eta}\bar{\lambda}=
-i \sqrt{2} {\cal D}_{\mu}\bar{A}\eta\sigma^{\mu}
+\sqrt{2}\bar{F}\bar{\eta}, 
\nonumber\\
\delta_{\eta}F&=&i\sqrt{2}\bar{\eta}\bar{\sigma}^{\mu}{\cal 
D}_{\mu}\lambda
- 2ig [A,\bar{\eta}\bar{\psi}],
\quad
\delta_{\eta}\bar{F}=-i\sqrt{2}{\cal D}_{\mu}\bar{\lambda}
\bar{\sigma}^{\mu}\eta
- 2ig [\eta\psi ,\bar{A}], 
\nonumber\\
\delta_{\eta}A_{\mu}&=&
i\bar{\psi} \bar{\sigma}_{\mu} \eta
-i\bar{\eta}\bar{\sigma}_{\mu}\psi, 
\nonumber\\
\delta_{\eta}\psi&=&
-\sigma^{\mu \nu}\eta F_{\mu \nu}-i\eta D, 
\quad \delta_{\eta}\bar{\psi}=
\bar{\eta} \bar{\sigma}^{\mu \nu} F_{\mu \nu}+i\bar{\eta} D, 
\nonumber\\
\delta_{\eta}D&=& \eta \sigma^{\mu}{\cal D}_{\mu}\bar{\psi}
+{\cal D}_{\mu}\psi\sigma^{\mu}\bar{\eta}.
\label{eq:n2susy2}
\end{eqnarray}

As in the ${\cal N}=1$ case, this theory has ${\cal N}={1\over2}$ 
supersymmetry, under which the component fields transform as 
\begin{eqnarray}
 \delta_{\xi}A&=& \sqrt{2}\xi\psi,\quad
\delta_{\xi}\bar{A}=0, 
\nonumber\\
\delta_{\xi}\psi_{\alpha}&=&
\sqrt{2}\xi_{\alpha} F,\quad
\delta_{\xi}\bar{\psi}_{\dot{\alpha}}
=-i\sqrt{2}{\cal D}_{\mu}\bar{A}(\xi\sigma^{\mu})_{\dot{\alpha}}, 
\nonumber\\
\delta_{\xi}F&=&0,\quad
\delta_{\xi}\bar{F}=
-i\sqrt{2}{\cal D}_{\mu}\bar{\psi}\bar{\sigma}^{\mu} \xi
+i [\xi\lambda,\bar{A}]
+C^{\mu\nu} \left( \partial_{\mu} \{\bar{A},\xi \sigma_{\nu} 
\bar{\lambda} \} 
+\frac{i}{2}[A_{\mu},\{\bar{A},\xi \sigma_{\nu} \bar{\lambda} \} ] 
\right) , \nonumber\\
\delta_{\xi}A_{\mu}&=&
-i\bar{\lambda}\bar{\sigma}_{\mu}\xi, 
\nonumber\\
\delta_{\xi}\lambda_{\alpha}&=& i\xi_{\alpha} D+
(\sigma^{\mu\nu}\xi)_{\alpha}\left(F_{\mu\nu}+{i\over2} C_{\mu\nu} 
\bar{\lambda}\bar{\lambda}\right)
,\quad
\delta_{\xi}\bar{\lambda}_{\dot{\alpha}}=0, 
\nonumber\\
\delta_{\xi}D&=& -\xi \sigma^{\mu}{\cal D}_{\mu}\bar{\lambda}. 
\end{eqnarray}
The transformations associated with $\bar{\xi}$ and
$\bar{\eta}$ are not the symmetry of the model.

One may ask whether the other symmetry parameterized by $\eta$ is
preserved or not.
Let us examine this symmetry 
in the case of ${\cal N}=2$ $U(1)$ gauge theory as an illustration.
The Lagrangian is 
\begin{eqnarray}
 {\cal L}^{{\cal N}=2}&=&
-{1\over 4}F^{\mu\nu}F_{\mu\nu}- 
i\bar{\lambda}\bar{\sigma}^{\mu}\partial_{\mu}\lambda
+{1\over2}D^2
\nonumber\\
&&-\partial^{\mu}\bar{A}\partial_{\mu}A
-i\bar{\psi}\bar{\sigma}^{\mu}\partial_{\mu}\psi+\bar{F}F
-\frac{i}{2} C^{\mu\nu}F_{\mu\nu}\bar{\lambda}\bar{\lambda}
+ i C^{\mu\nu}F_{\mu\nu} \bar{A} F , 
\label{eq:U(1) lagrang}
\end{eqnarray}
and the transformation (\ref{eq:n2susy2}) for $\bar{\eta}=0$
becomes
\begin{eqnarray}
\delta_{\eta}^0A&=&\sqrt{2}\eta\lambda,\quad \delta_{\eta}^0\bar{A}=0,
\nonumber\\
\delta_{\eta}^0\lambda&=& \sqrt{2}\eta F,\quad
\delta_{\eta}^0\bar{\lambda}=-\sqrt{2}\partial_{\mu}\bar{A}
\eta\sigma^{\mu},\nonumber\\
\delta_{\eta}^0F&=&0,\quad
\delta_{\eta}^0\bar{F}=-i\sqrt{2}\partial_{\mu}\bar{\lambda}
\bar{\sigma}^{\mu}\eta \nonumber\\
\delta_{\eta}^0 A_{\mu}&=& i\bar{\psi}\bar{\sigma}_{\mu}\eta, 
\nonumber\\
\delta_{\eta}^0\psi&=& -\sigma^{\mu\nu}\eta F_{\mu\nu},\quad
\delta_{\eta}^0\bar{\psi}=0, 
\nonumber\\
\delta_{\eta}^0D&=&\eta \sigma^{\mu}\partial_{\mu}\bar{\psi}. 
\label{eq:n2}
\end{eqnarray}
One may expect that this theory does not have any deformed 
$\eta$-symmetry, 
because the $C$-dependent term in the Lagrangian does not have  
the $SU(2)_{R}$ ${\cal R}$ symmetry 
(under which $(\psi,\lambda)$ transform as the doublet). 
This is indeed the case as we will see below. 
The variation of the last term in (\ref{eq:U(1) lagrang}) 
can be cancelled by adding a $C$-dependent term to 
the transformation of $\bar{F}$. 
The transformation of the remaining $C$-dependent term 
in the Lagrangian is 
\begin{equation}
\delta_{\eta}^0 (-\frac{i}{2} C^{\mu\nu}F_{\mu\nu}\bar{\lambda} 
 \bar{\lambda})
= C^{\mu \nu} \partial_{\mu}\bar{\psi}\bar{\sigma}_{\nu}\eta 
\bar{\lambda}\bar{\lambda}
+ \sqrt{2}i C^{\mu\nu}F_{\mu\nu}\partial_{\rho}\bar{A} \eta 
\sigma^{\rho}\bar{\lambda}.
\label{eq:c1}
\end{equation}
These terms must be cancelled by adding $C$-dependent
terms to the transformation (\ref{eq:n2}).
But it is shown that the second term cannot be cancelled 
by any $C$-dependent deformation.
In the Lagrangian, only the terms 
$\bar{\lambda}\bar{\sigma}^{\mu}\partial_{\mu}\lambda$,  
$\partial^{\mu}\bar{A}\partial_{\mu}A$ 
or $F^{\mu\nu}F_{\mu\nu}$ could give 
the variation of the form 
$C^{\rho\sigma}F_{\rho\sigma}\partial_{\mu}\bar{A}
\eta\sigma^{\mu}\bar{\lambda}$. 
In the first case, 
$\delta_{\eta}^C (\partial_{\mu}\lambda) \propto 
\eta C^{\rho\sigma}F_{\rho\sigma}\partial_{\mu}\bar{A}$ should hold. 
But this means that $\delta_{\eta}^C\lambda$ should be an integration of 
$\eta C^{\rho\sigma}F_{\rho\sigma}\partial_{\mu}\bar{A}$, 
which cannot be a local transformation. 
Thus in this case, so far as the local transformations, 
it is impossible to construct a deformed $\eta$-transformation. 
The other two possibilities are also excluded with much the same argument. 
The above argument can be extended to the case of $U(N)$ gauge groups.
Under the transformation parameterized by $\eta$, 
it is found that the variations of 
the $C$-deformed part of the Lagrangian 
contains the only one term proportional to $D$: 
$C^{\mu \nu} [{\cal D}_{\mu} \bar{A} , \bar{\lambda}] \bar{\sigma}_{\nu} 
\eta D$. 
The argument similar to the $U(1)$ case shows that 
this term cannot be cancelled 
by any $C$-deformation of the $\eta$-transformation 
so far as the local transformations. 

We have constructed the gauge invariant Lagrangian 
for ${\cal N}=1$ supersymmetric gauge theory 
on the noncommutative superspace. 
We have shown that it is possible to construct 
the $C$-deformed supersymmetry transformation generated by $Q_{\alpha}$. 
This means that 
the theory has ${\cal N}=\frac{1}{2}$ supersymmetry 
as in the case of \cite{Se}. 
Generalizing the argument, 
we have seen that ${\cal N}=2$ supersymmetric Yang-Mills theory 
on the noncommutative superspace has only ${\cal N}={1\over2}$ supersymmetry.
This may be seen from the fact that 
the $C$-deformed terms are not invariant under
the $SU(2)_{R}$ ${\cal R}$ symmetry.
In view of extended supersymmetry, it would be interesting to
investigate the deformation of ${\cal N}=2$ rigid superspace\cite{GrSoWe}.
If one may deform the superspace with ${\cal R}$ symmetry, it would
be possible to construct the noncommutative supersymmetric gauge theory
with extended supersymmetry.

 It is interesting to study non-perturbative aspects of the
$C$-deformed theory by studying  solitons, instantons and monopoles etc.
It would be also interesting to study 
whether such deformations are possible in three or two dimensions.

\end{document}